\documentclass[twoside,twocolumn]{elsart3}

\usepackage{amssymb}
\usepackage{amsmath}
\usepackage{latexsym}
\ifx\pdftexversion\undefined
  \usepackage[dvips]{graphics}
\else
  \usepackage[pdftex]{graphics}
\fi
\usepackage{graphicx} 
\usepackage{tabularx}

\begin{document}

\def\um{\,\mu{\rm  m}}

\def\enc   {\ensuremath{\mathrm{ENC}}\xspace}
\def\Ileak {\ensuremath{I_\mathrm{{leak}}}}

\def\nm{\mathrm{\,   nm}}
\def\um{\mathrm{\,\mu m}}
\def\mm{\mathrm{\,   mm}}
\def\cm{\mathrm{\,   cm}}
\def \m{\mathrm{\,    m}}

\def\fs{\mathrm{\,   fs}}
\def\ps{\mathrm{\,   ps}}
\def\ns{\mathrm{\,   ns}}
\def\us{\mathrm{\,\mu s}}
\def\ms{\mathrm{\,   ms}}

\def\mOhm{\mathrm{\,m\Omega}}
\def\Ohm {\mathrm{\, \Omega}}
\def\kOhm{\mathrm{\,k\Omega}}
\def\MOhm{\mathrm{\,M\Omega}}
\def\GOhm{\mathrm{\,G\Omega}}
\def\TOhm{\mathrm{\,T\Omega}}

\def\nV  {\mathrm{\,  n V}}
\def\uV  {\mathrm{\,\mu V}}
\def\mV  {\mathrm{\,  m V}}
\def \V  {\mathrm{\,    V}}

\def\aA  {\mathrm{\,  a A}}
\def\fA  {\mathrm{\,  f A}}
\def\pA  {\mathrm{\,  p A}}
\def\nA  {\mathrm{\,  n A}}
\def\uA  {\mathrm{\,\mu A}}
\def\mA  {\mathrm{\,  m A}}
\def\Amp {\mathrm{\,    A}}

\def\nS  {\mathrm{\,  n S}}
\def\uS  {\mathrm{\,\mu S}}
\def\mS  {\mathrm{\,  m S}}

\def\aF  {\mathrm{\,  a F}}
\def\fF  {\mathrm{\,  f F}}
\def\pF  {\mathrm{\,  p F}}
\def\nF  {\mathrm{\,  n F}}
\def\uF  {\mathrm{\,\mu F}}
\def\mF  {\mathrm{\,  m F}}
\def \F  {\mathrm{\,    F}}

\def\nH  {\mathrm{\,n   H}}
\def\uH  {\mathrm{\,\mu H}}

\def\pC  {\mathrm{\,  p C}}
\def\fC  {\mathrm{\,  f C}}

\def\mHz{\mathrm{\,mHz}}
\def\Hz {\mathrm{\, Hz}}
\def\kHz{\mathrm{\,kHz}}
\def\MHz{\mathrm{\,MHz}}
\def\GHz{\mathrm{\,GHz}}
\def\THz{\mathrm{\,THz}}

\def\eV {\mathrm{\,eV}}
\def\keV{\mathrm{\,keV}}
\def\GeV{\mathrm{\,GeV}}

\def\uW {\mathrm{\,\mu W}}
\def\mW {\mathrm{\,  m W}}
\def\W  {\mathrm{\,    W}}

\def\bit   {\,\mathrm{bit}}
\def\kbit  {\,\mathrm{kbit}}

\def\eps   {{\epsilon}}
\def\epsO  {{\epsilon_0}}
\def\epsSi {{\epsilon_{Si}}}
\def\epsSiO{{\epsilon_{SiO_2}}}

\def\electrons {\mathrm{\,e^-}}
\def\e  {\mathrm{\,e^-}}
\def\half{{1\over 2}}

\def\kbit{\,\mathrm{kbit}}
\def\bit {\,\mathrm{bit}}

\renewcommand{\labelenumi}{\arabic{enumi}}
\renewcommand{\labelitemi}{-}

\begin{frontmatter}

\title{Concept, realization and characterization of serially powered pixel modules (Serial Powering)}


\author[bonn]{D.B.~Ta},
\author[bonn]{T.~Stockmanns}
\author[bonn]{F.~H\"ugging}
\author[bonn]{P.~Fischer}
\author[bonn]{J.~Grosse-Knetter},
\author[bonn]{\"O.~Runolfsson},
\author[bonn]{N.~Wermes}


\address[bonn]{Physikalisches Institut der Universit{\"a}t Bonn\\ Nussallee 12,                D-53115 Bonn, Germany\\                 Tel.: +49/228/73-2352, Fax:                -3220\\ email: ta@physik.uni-bonn.de}

\begin{abstract}
We prove and demonstrate here for the example of the large scale pixel detector of ATLAS that {\sl Serial Powering} of pixel modules is a viable alternative and that has been devised and implemented for ATLAS pixel modules using dedicated on-chip voltage regulators and modified flex hybrids circuits. The equivalent of a pixel ladder consisting of six serially powered pixel modules with about 0.3\,Mpixels has been built and the performance with respect to noise and threshold stability and operation failures has been studied. We believe that {\sl Serial Powering} in general will be necessary for future large scale tracking detectors.
\end{abstract}


\end{frontmatter}

\newpage

\section{Introduction}
\label{sec:introduction}
Modern particle detectors require on the one hand a large solid angle coverage and high granularity, on the other hand fast read-out and low power consumption. Especially particle trackers require in addition a minimum of passive material inside their active region. The high granularity results in building a detector from a large number of identical active modules. The usual power scheme is the individual, parallel powering of the modules with a constant voltage. However, this is disadvantageous for a large scale detector such as a pixel detector.

The pixel detector as the innermost sub-detector of many large scale particle detectors has a very high granularity. The ATLAS pixel detector comprises 1744 active pixel modules containing about 80 million channels~\cite{atlaspixeltdr}. It makes use of deep sub-micron (0.25$\um$) chip technology for the read-out which is necessary to achieve a radiation tolerant, compact, and high granularity detector design. Each hybrid module is composed of a 250$\um$ thick silicon sensor and 16 Front-End chips with a total of approx. 46\,000 pixels with a size of $400\times 50\um$.

In order to achieve a fast operation of the detector, the electronic circuits must be powered with high currents (approx.~$2\Amp$) at low voltages (1.6$\V$ analog, 2.0$\V$ digital). Therefore a high power density of the detector comes with the high granularity. Since the power supplies are located outside the active detector volume, the power is transmitted over a long distance (which can easily attain $100\m$) and the voltages are regulated remotely from outside the active volume of the detector (for a large scale detector the distance can be larger than 10$\m$). Further, the power cable diameter is small inside the detector volume (typically $300\um$ diameter Al wires). Together with the low voltages and high currents the power losses in the cables exceed the actual power consumption of the modules. Moreover, the further the granularity increases, the more cables are needed. This results not only in an increase of the power losses, but also increases the passive material in the active detector volume. Hence, from a certain granularity onwards the parallel powering concept has more and more disadvantages and it will become more favorable to change to the new serial powering scheme.

\begin{figure}
  \centering
  \includegraphics[width=0.9\linewidth]{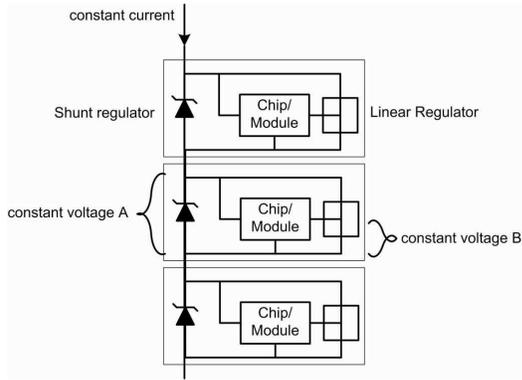}
  \caption{Serial Powering Scheme, in which the voltages for the chips/modules are produced from a constant current by shunt and linear regulators. Only two power cables are needed to power a chain of modules, instead of four (eight, if sense lines are included) power cables for each module.}
  \label{fig:simplechain}
\end{figure}

For the {\sl Serial Powering} concept~\cite{tobiaspaper} a chain of modules is powered in series by a constant current. Only two power lines per chain are needed. The voltages are generated by shunt and linear regulators implemented on the chip itself (see figure~\ref{fig:simplechain}). {\sl Serial Powering} offers a drastic reduction of power lines and therefore a reduction in passive materials in the active detector region as well as a reduction of power losses in the cables. A positive side effect is the reduction of the heat from the power lines that can uncontrollably heat up other detector systems.

In this paper we demonstrate for the ATLAS Pixel Detector that {\sl Serial Powering} is a viable alternative for a possible upgrade of the ATLAS Inner Detector and that all the possible problems listed in the next section, concerning stability and reliability, can be solved if a dedicated {\sl Serial Powering} design is used.

\section{Example calculation on the ATLAS pixel detector}

Every module is connected to 8 power cables, i.e. two power lines and two sense lines for every voltage. The total voltage drop from the power supplies in the control room to the module is 6.4$\V$ and the power losses in the cables are $191\W$ maximum. This is $\sim 2.81$ times of the power consumption of the 13 modules. The total lengths of all cables in the active region are $121\m$, this corresponds to an average radiation length of 0.073\,\%$X_0$.

Since the modules of the pixel detector are organized in 114 pixel ladders (staves) with 13 modules, this suggests that the 13 modules of such a stave form a chain of serially powered modules.
The advantages of a serially powered chain of 13 modules follow from an example calculation (table~\ref{tab:comparision_pp_ss}):
\begin{itemize}
\item {Passive material, i.e. power cables in the detector is reduced by a factor of 50 compared to the Parallel Powering scheme. This corresponds to 15\% of the radiation length of the parallel scheme.}
\item {Power losses in the cables are reduced by a factor 10.}
\item {Reduced power losses in the cables also reduce heat-pick up by other detector systems.}
\item {Voltage regulation is done on-chip, i.e. close to the consumer. The regulation can respond faster to voltage changes due to varying power consumptions.}
\end{itemize}

\begin{table}

    \caption{Comparison between {\sl Parallel} (using numbers from~\cite{atlipes0007}) and {\sl Serial Powering}, the power consumption is given for a ladder of 13 modules. The different {\sl Serial Powering} schemes differ in the voltage that drops across a module.}
    \label{tab:comparision_pp_ss}
   \begin{tabularx}{\linewidth}{Xb{1cm}b{1cm}b{1cm}}
        \hline
& Parallel & Serial & Factor\\ \hline
        \hline
Power supplies &  13 & 1 & 13 \\
No. of power lines &104 & 2 & 52 \\
Total cable length (mm) &   121160  & 2780 & 43.6 \\
        \hline
    \hline
radiation length per layer $x/X_0$ & 0.073\,\% & 0.011\,\% & 6.5  \\
        \hline
Cable power losses&  191\,W &  19.2\,W &  9.97 \\
Module (Shuntreg. 2.0\,V) &    67.9\,W &  78\,W & 0.87\\
Module (Shuntreg. 2.7\,V) &   & 105.3\,W & 0.65 \\ \hline
Sum (Shuntreg. 2.0\,V) &  259\,W &  97.2\,W &   2.67 \\
Sum (Shuntreg. 2.7\,V) & &  124.5\,W    & 2.08  \\ \hline
      \end{tabularx}
\end{table}

Apparently also several concerns about {\sl Serial Powering} arise. The first concern that the local power consumption of the module is higher than for parallel powered modules can be rejected partially from a separate calculation which includes the total heat load for the cooling system. The heat load from the cables\footnote{The heat load per cable is \hbox{approx.~$2\W$}.} that are closest to the module is transferred to the modules and the decrease of the heat load due to less cables counterbalances the increase of the module heat.  The heat that must be cooled by the cooling system decreases by 7\% ({\sl Serial Powering} scheme with 2.0$\,V$ voltage drop across a module) or increases maximally by 17\% (scheme with 2.7$\,V$ voltage drop across a module). Additionally, the heat is now produced close to the cooling and as mentioned above there is less heat pickup by other detector systems, so that the slight increase of the heat load has overall positive effects on the detector.

The fear of loss of a whole chain due to one defect regulator is addressed by the {\sl Serial Powering} scheme (see figure~\ref{sec:general_sp_scheme}). The major concern of noise pickup of the chain by noise sources or noisy modules is rejected by the measurements in~\ref{subsec:failuremodes}.


\section{General Serial Powering Scheme}
\begin{figure}
  \centering
  \includegraphics[width=0.9\linewidth]{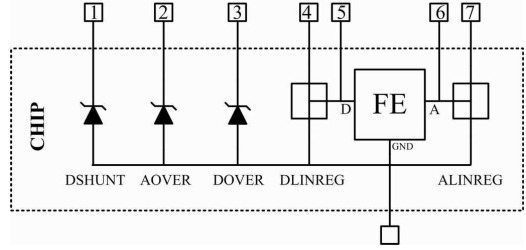}
  \caption{Schematic of the powering components of a chip. The analog/digital circuits of the chip is indicated by FE. The
different powering schemes are realized by proper connections
between the input pads (pads 5 and 6 are only necessary for
\sl{Parallel Powering}).}
  \label{fig:fechip}
\end{figure}

There are three shunt regulators on the ATLAS pixel production
chip, DSHUNT, AOVER and DOVER that have design threshold voltages
of 2.0$\V$, 2.4$\V$, and 2.7$\V$, respectively. The two linear
regulators ALINREG and DLINREG have adjustable output voltages in
four steps from 1.5$\V$ to 1.8$\V$ or 1.8$\V$ to 2.4$\V$, resp.
Both regulators are described in detail
in~\cite{tobiaspaper}~\cite{ivandoktorarbeit} and ~\cite{dbt}. The
different powering schemes are realized by proper connections
between the input pads of the different regulators on the chip
(figure~\ref{fig:fechip} show the schematically the powering
components of a chip).

\begin{figure}
  \centering
  \includegraphics[width=0.9\linewidth]{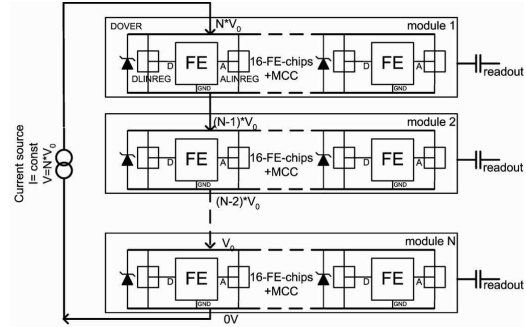}
  \caption{A chain of serially powered modules, each built according to the {\sl Extended Serial Powering} scheme}\label{fig:chainof_sp_modules}
\end{figure}

Figure~\ref{fig:chainof_sp_modules} shows a chain of serially
powered modules ({\sl Extended Serial Powering}
scheme~\cite{tobiaspaper}, by connecting pad 3, 4 and 7 in figure
\ref{fig:fechip} to the constant current source). A constant
current source is connected to the power input of the first
module. The ground of the module is then connected to the power
input of the next module etc. The current consumption is
determined by the highest current consumption of one module in the
chain. The supply voltages of the chips are generated by the two
types of regulators. One is a shunt regulator which behaves like a
Zener diode. The other one is a linear voltage regulator.

\label{sec:general_sp_scheme}

On a module the shunt regulators of all 16 chips are connected in
parallel. This stabilizes the output voltage. The fear of a loss
of a whole module chain is also addressed this way. The redundant
use of the shunt regulators can maintain the voltage regulation
and the chain remains uninterrupted even if one or more regulators
should break (i.e. creates an open-circuit). Thus the risk of
loosing a whole chain is minimized. The common output voltage of
the shunt regulators is the input voltage for the two linear
regulators of every chip. The outputs of the regulators are
already internally connected to the analog and digital part,
respectively. The suitable regulators for this scheme are
therefore the two linear regulators that are powered by the DOVER
regulator.
As the modules now have different ground potentials, the read-out must be done via AC coupling of the signals (as indicated in figure~\ref{fig:chainof_sp_modules} by a symbolic capacitor).


\begin{figure}
  \centering
  \includegraphics[width=0.75\linewidth]{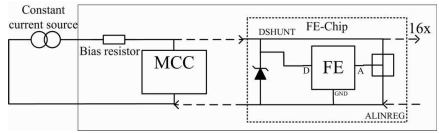}
  \caption{Schematic of the {\sl Simple Serial Powering} scheme using only one linear regulator (only one single serially powered module, further modules are added as shown in figure~\ref{fig:chainof_sp_modules}).}\label{fig:sp-simple}
\end{figure}

One alternative {\sl Serial Powering} schemes can also be considered. The {\sl Simple Serial Powering} scheme (fig.~\ref{fig:sp-simple} by connecting pad 1, 5 and 7 to the constant current source) uses the DSHUNT regulator to power the digital part and the analog linear regulator. In this scheme the total voltage which drops across a module is lowest (2.0$\V$) and therefore also the power consumption of a single module is the lowest. A disadvantage is that the digital voltage is fixed to the threshold voltage of the shunt regulator.

\section{Operation of a serially powered module ladder}
\label{sec:stave}
\begin{figure}
  \centering
  \includegraphics[width=0.9\linewidth]{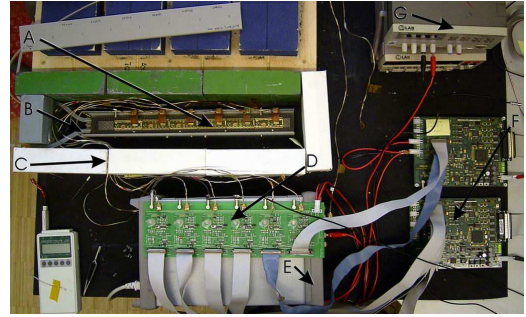}
  \caption{Picture of the measurement setup with (A) the module ladder (half-stave) with six serially powered modules containing about 0.3\,Mpixels, (B) cooling pipes, (C) power cables, (D) AC coupling read-out and power routing board, (E) constant current power supply, (F) pixel read-out electronics, (G) power supplies for read-out electronics}\label{fig:stave}
\end{figure}

\begin{figure}

  \includegraphics[width=0.9\linewidth]{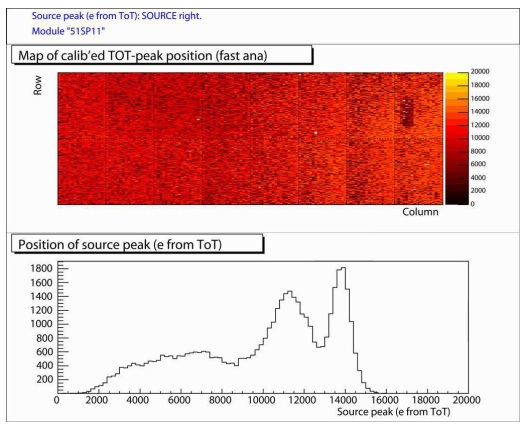}
\caption{Spectrum of a ${}^{241}Am$ source placed between two
modules and recorded simultaneously (picture shows only the left
module)}
\label{fig:left}
\end{figure}

After testing and comparing single serially powered modules~\cite{tobiaspaper}~\cite{dbt}, which were built with already existing ATLAS pixel chips and a dedicated flex-hybrid,  six {\sl Serial Powering} modules were composed to a ladder, a so called half-stave, containing about 0.3\,$\mathrm{MPixels}$. The measurement setup is shown in figure~\ref{fig:stave}. The same original mechanical support structure for module ladders was used except for the reduced number of modules. The serial connections between the modules was made by original power cables (type-0) which were connected to a special board that routes the current serially through all modules. Additionally it AC-couples the LVDS signals between the modules and the external read-out electronics through a pair of capacitors and a LVDS buffer for each data line. 
The operation commands, i.e. commands that injects test charges
into the pixels and initiates the read-out chain, could be sent
to all modules at the same time. This emulates the situation in
the detector when all modules are individually having digital and
analog activities and varying power consumptions. As first
qualitative proof of operation, figure~\ref{fig:left} shows the
spectrum of a ${}^{241}\mathrm{Am}$ point source placed above and
between two modules and recorded simultaneously. Only shown is the
left hit map which corresponds to the module that was on the left
side of the source. In the spectrum two peaks can be seen. The
peak at higher deposited charge is due to the absorbed
59.54$\keV$-photon\footnote{The second peak results from gamma
rays that transverse the sensor in an shallow angle so that
charge is shared among two pixels. In one of the pixels the
charge fraction is below the threshold. This results in a second
peak which is shifted by threshold.}. This tests the whole
detection chain from the charge collection in the sensor, the hit
processing in the chip to the output of the hit data. It is a
qualitative proof of the full module functionality.

\subsection{Performance Characterization}
\label{subsec:staveperformance} 
\begin{figure}
  \centering
  \includegraphics[width=0.9\linewidth]{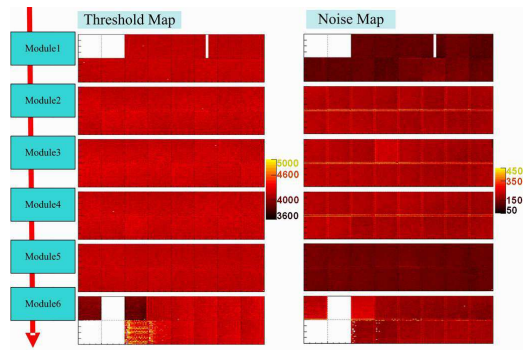}
  \caption{Thresholds and noise maps of the six serially powered modules (some modules without sensor or with defect chips) on the module ladder (half-stave)}\label{fig:all6maps}
\end{figure}

The performance of each module
while serially powered with five other actively working modules
was measured (multi-module operation mode). The measurements were
repeated until all six modules were read out.
Figure~\ref{fig:all6maps} shows the threshold maps and the noise
maps of the six modules\footnote{Some modules are without sensor
or with some defect chips because of cost reasons, still the vast
amount of pixels can give a good impression of the module's
performance.} and table~\ref{tab:staveperformance} shows the
threshold (THRES), threshold dispersion ($\sigma_\mathrm{THRES}$)
and the noise performance of the six modules. The noise
difference between the single serially powered operation (only
one single active, serially powered module) and the multi-module
operation mode is given in brackets. This is the first time that
such a large chain of serially powered modules was powered and
that all modules were operated at the same time. The results in
comparison to singly serially powered operation and parallel
powered modules show hardly any influence of the modules on each
other during normal operations.

\begin{table}
\caption{Threshold, threshold dispersion and noise of six
serially powered modules on a module ladder (in brackets: noise
difference to singly serially powered). Except for module 6 all
modules were built according to the {\sl Extended Serial Powering
scheme}.} \label{tab:staveperformance}
 \begin{tabularx}{\linewidth}{X b{1.75cm} b{1.75cm} b{1.75cm}}
 \hline
 Module & THRES ($e^-$) & $\sigma_\mathrm{THRES}$ ($e^-$) & noise \,\,\, ($e^-$)\\ \hline
        M1 & 4134  & 57      & $127^+$ (4)\\
        M2 & 4156  & 69      & 182 (-1)\\
        M3 & 4173  & 70      & 186 (-0)\\
        M4 & 4162  & 70      & 183 (-4)\\
        M5 & 4132  & 58      & $133^+$ (0)\\
        M6 & 4160  & 91      & 172 (-5)\\\hline
         {\sl Parallel}   &  4062     & 50 & 160      \\ \hline
\multicolumn{2}{l}{$^+$ \footnotesize without sensor}\\
\hline
 \end{tabularx}
\end{table}

\subsection{Failure Mode Studies}
\label{subsec:failuremodes} Two types of failures were studied in
order to test the reliability of the {\sl Serial Powering}
scheme. The first was the study of the effect of a noisy module
in the chain. This was achieved by lowering the detection
threshold of one module to almost zero, so that this module sees
noise hits all the time. That involves a permanent analog and
digital activity as well as permanent changing current
consumptions which can modulate the powering of the other modules
by permanently changing the potentials across the modules. The
measurement was repeated six times, so that each time a different
module was made noisy and the effect on the other modules was
investigated. Table~\ref{tab:GDAC0a} shows the noise difference
of every module between the failure mode measurement and the
normal operation. Each row is a different measurement with a
different noisy module.

\begin{table}
    \caption{Noise difference of the modules on the module ladder between normal operation and operation with one noisy module in the chain. Each row is a different measurement with a different noise module. Except for module 6 all modules were built according to the {\sl Extended Serial Powering scheme}.}
    \label{tab:GDAC0a}
\begin{tabularx}{\linewidth}{X b{0.8cm} b{0.8cm} b{0.8cm} b{0.8cm} b{0.8cm} b{0.8cm}}
\hline
Noisy module & M1 ($e^-$) &    M2 ($e^-$)&    M3 ($e^-$)&    M4 ($e^-$) &    M5 ($e^-$)&  M6 ($e^-$)\\
\hline
Module 1 &      --- &     13 &     2 &       2 &      13 &       3 \\
Module 2 &      6 &       --- &    10 &      5 &      0 &       9 \\
Module 3 &      0 &       2 &      --- &     2 &      0 &       3 \\
Module 4 &      1 &       9 &      2 &       --- &    0 &       9 \\
Module 5 &      10 &      2 &      1 &      15 &      --- &     20 \\
Module 6 &      0 &       2 &      2 &       2 &      0 &       --- \\
\hline
\end{tabularx}
\end{table}

The increase in noise for neighboring modules in the chain is small (max. $20\e$) compared to the normal noise of $170\e$ to $200\e$ and it is well below the ATLAS requirement for the maximum noise. This shows that the filtering and the regulation of the supply voltages is sufficient enough to filter such a disturbance on the power lines.


The second type of failure mode tested is a noisy module emulated by an external switchable load connected in parallel to one module. The distortion was tuned to different frequencies up to 40$\MHz$. The load varied between 300$\mA$ and 500$\mA$. This constitutes a much more massive interference to the supply current than a noisy module. The effect on the chain is again a shift in potentials for all modules.

\begin{figure}

  \includegraphics[width=0.9\linewidth]{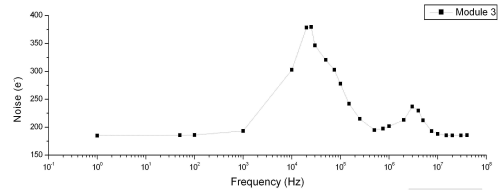}
 \caption{Frequency dependence of the noise of the module that was parallel to the switchable load}
  \label{fig:fsingle}
\end{figure}

\begin{figure}

  \includegraphics[width=0.9\linewidth]{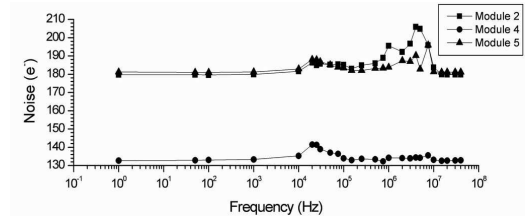}
\caption{Frequency dependence of the noise of the modules
neighboring the module that was parallel to the load}
  \label{fig:fall}
\end{figure}

Figures~\ref{fig:fsingle} and \ref{fig:fall} shows the frequency
dependence of the noise of  module~3 which was parallel to the
switchable load. During the whole measurement module~3 could be
configured and read out sequentially without any problems. Only in
two regions around 25$\kHz$ and around 3$\MHz$ the noise rises to
undesired values of 350$\e$ resp. 230$\e$. More important is to
observe the effect on the other modules, whether neighboring
modules pick up the higher noise from module 3.
Figure~\ref{fig:fall} shows the frequency dependence of the noise
of the three neighboring modules in the chain. Clearly the change
in noise is small for all modules\footnote{module 5 is again
without sensor}. Therefore there is hardly any noise pick up by
{\sl Serial Powering} modules from noise on the power line. We
consider this experimental test a very important demonstration
that the {\sl Serial Powering} scheme is reliable.


\section{Summary}
We have demonstrated that {\sl Serial Powering} is a viable concept to provide
operation power to a large pixel detector system such as the ATLAS
pixel detector at LHC. A powering scheme that powers a chain of modules with a constant current and uses dedicated on-chip voltage regulators and modified flex hybrid circuits has been devised and implemented for ATLAS pixel modules.

An example calculation shows that such a chain of 13 modules offers a reduction in power losses of the cables by 90\% and a reduction in passive materials by 98\%, this is a reduction of 85\% in radiation length.
It has been shown that the spread in quality of the voltage
regulators, as the key elements to this powering scheme, is
sufficiently small and the voltage stability of the linear
regulators is excellent, so that the voltage regulators are
applicable for Serial Powering. The serially powered modules have
been intensively tested in the lab. The comparison between
parallelly powered and serially powered modules has shown no
difference between the two powering schemes. Finally, the
equivalent of a pixel ladder consisting of six serially powered
pixel modules with about 0.3\,Mpixels has been built and the
performance with respect to operation failures has been studied.
Measurements with artificially noisy modules mimicked by inducing
noise on the power lines have only shown a marginal increase in
noise of the other modules in the chain. We strongly believe that
Serial Powering is not only a viable powering scheme for an
upcoming upgrade of the ATLAS pixel detector, but is also viable,
if not necessary for future large scale tracking detectors.




The thesis of the university of Bonn can be found under
\verb|hep1.physik.uni-bonn.de| under the section
\verb|Publications|.

This paper is a short version of the following paper:

D.B. Ta et~al.
\newblock {\em Serial Powering: Proof of principle demonstration of a scheme for the operation of a large pixel detector at the LHC}
\newblock Nucl. Instr. \& Meth. A557 (2006) 445--459.

\end{document}